\documentclass{appolb}
\usepackage{epsfig}

\begin{document}
\pagestyle{plain}
\newcount\eLiNe\eLiNe=\inputlineno\advance\eLiNe by -1
\title{RECENT PROGRESS AND PUZZLES IN CHARMONIUM PHYSICS}
\author{B. G. FULSOM
\address{University of British Columbia,
Vancouver, British Columbia, Canada, V6T 1Z1}}
\maketitle

\begin{abstract}
While the charmonium model has been effective in describing $c\overline{c}$ bound mesons,
there have been many recently discovered charmonium-like states it cannot accommodate.
Here I provide a review of recent results from the $B$-factories including the $X(3872)$,
three new particles in the mass range near $3.93$ GeV$/c^{2}$, and four new resonances in initial
state radiation (ISR) decays.
\end{abstract}

\section{Introduction}
\begin{figure}
\begin{center}
\epsfig{file={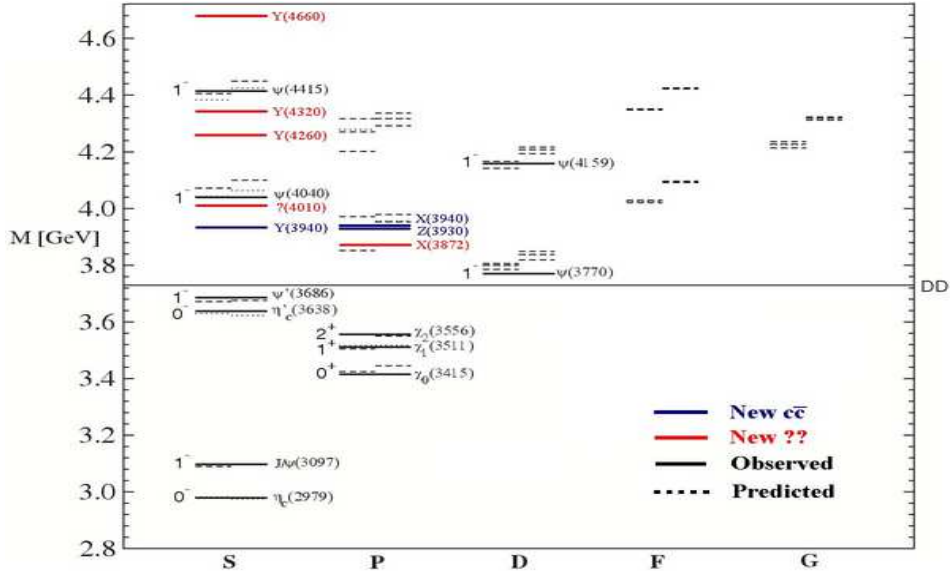},width={5.0in},height={3.0in}}
\caption{The charmonium spectrum \cite{figure}.} \label{fig_1}
\end{center}
\end{figure}

The charmonium model is a phenomenological model describing the bound state of the charm and anti-charm quark system \cite{theory}. Figure \ref{fig_1} demonstrates the correspondence between experiment and theory of the charmonium spectrum for two selected models \cite{figure}. The dashed lines illustrate theoretically predicted masses, overlaid with black solid lines indicating the well-established experimental results, and red and blue solid lines for recently discovered resonances yet to be incorporated into the model. In the case of the well-established states, there is very good agreement between the theory and experiment. The series of newly discovered charmonium-like states will be the primary focus of this talk.

I will concentrate on results obtained at the BABAR and Belle $B$-factories. The BABAR results are based on $200-350$ fb$^{-1}$ of $e^{+}e^{-}$ collisions at the $\Upsilon(4S)$ resonance ($\sqrt{s}\approx10.58$ GeV) at the PEP-II linear accelerator at SLAC. The Belle results are from up to $\approx700$ fb$^{-1}$ of the same type of collisions at the KEK-B accelerator at KEK.

\section{$X(3872)$}
In 2003, Belle discovered a signal in the decay $B^{+}\rightarrow X K^{+}$, $X\rightarrow J/\psi\pi^{+}\pi^{-}$ \cite{x3872_belle}. This state was found to have a mass of $m_{X}=3871.2\pm0.6$ MeV$/c^{2}$ and a very narrow width, $\Gamma<2.3$ MeV. This discovery was later verified by CDF, D0, and BABAR \cite{x3872_other}. Evidence for $X\rightarrow\gamma J\psi$ determines C-parity to be positive \cite{x3872_radiative}. Angular analyses from Belle and CDF \cite{x3872_angular} favour $J^{PC}=1^{++}$. No charged partners of the $X(3872)$ have been found, and decays to $\chi_{c1,2}\gamma$ and $J/\psi\eta$ have not been seen.

The $X(3872)$ displays some characteristics of a charmonium-like state, but its narrow width above the $D\overline{D}$ threshold, mass, and quantum numbers do not correspond with charmonium model predictions. It is important to consider $m_{X}\approx m_{D}+m_{\overline{D}^{*0}}$, leading to speculation that the $X(3872)$ may be the bound state of two $D$ mesons, i.e. a $D^{0}\overline{D}^{*0}$ molecule. This is supported by predictions for its mass, decay modes, $J^{PC}$ values, and branching fractions. Other more exotic interpretations include tetraquark, glueball, or charmonium-gluon hybrid bound states. For a summary of theoretical interpretations of the $X(3872)$, see \cite{x3872_theory}.

\section{$X(3940)$, $Y(3930)$, $Z(3940)$}
Belle has discovered three more charmonium-like states in a similar mass range via distinct production methods and decay modes. All three states have plausible charmonium model interpretations \cite{godfrey_fpcp}.

The $X(3940)$ was discovered by the recoil of the $J/\psi$ in the double-charmonium production of $e^{+}e^{-}\rightarrow J/\psi X(3940)$ on 350 fb$^{-1}$ of data \cite{x3940_belle}. It was found to decay to $DD^{*}$ but not $DD$. This points towards an assignment as the $\eta_{c}(3S)$.

The $Y(3930)$ was seen in the decay $B\rightarrow K Y$, $Y\rightarrow J/\psi\omega$. In Belle's dataset of 278M $B$ decays, they measured a mass and width of $m_{Y}=3943\pm11\pm13$ MeV$/c^{2}$ and $\Gamma(Y)=87\pm22\pm26$ MeV \cite{y3930_belle}. This state was confirmed by BABAR \cite{y3930_babar}, but using 385M $B$ decays they measured it to have a mass and width of $m_{Y}=3914.3^{+3.8}_{-3.4}\pm1.6$ MeV$/c^{2}$ and $\Gamma(Y)=33^{+12}_{-8}\pm1$ MeV. An apparent interpretation of the $Y(3930)$ state is the $\chi_{c1}(2P)$ charmonium state.

Finally, using 395 fb$^{-1}$ of data, the $Z(3930)$ was found by Belle in the two-photon process $\gamma\gamma\rightarrow Z(3930)$ and decaying to $D\overline{D}$ \cite{z3930_belle}. The $\chi_{c2}(2P)$ charmonium assignment is an eminent choice based on its production, decay, mass, and width.

\section{States produced in ISR}
Several new states have been discovered via initial-state-radiation production. The first of these was BABAR's discovery \cite{y4260_babar} of a broad structure in the decay $e^{+}e^{-}\rightarrow Y(4260)$, $Y(4260)\rightarrow J/\psi\pi^{+}\pi^{-}$. Based on 211 fb$^{-1}$ of data, the mass and width of this bump were measured to be $m_{Y}=4259\pm8^{+2}_{-6}9$ MeV$/c^{2}$ and $\Gamma(Y)=88\pm23^{+6}_{-4}$. Following this discovery, CLEO performed a centre-of-mass energy scan and collected data directly at the $Y(4260)$ resonance \cite{y4260_cleo}. Reconstructing 16 decay modes, they confirmed BABAR's discovery as well as finding evidence for $Y(4260)\rightarrow J/\psi\pi^{0}\pi^{0}$ and $Y(4260)\rightarrow J/\psi K^{+}K^{-}$. Using 550 fb$^{-1}$ of data, Belle has also confirmed BABAR's discovery \cite{y4260_belle}, measuring a mass of $m_{Y}=4247\pm12^{+17}_{-26}$ MeV$/c^{2}$ and a width of $\Gamma(Y)=108\pm19^{+8}_{-10}$ MeV. Additionally, Belle claims a second much broader resonance at $m=4008\pm40^{+72}_{-28}$ MeV$/c^{2}$ with a width of $\Gamma=226\pm44^{+87}_{-79}$ MeV. Because these states are produced in the annihilation of $e^{+}e^{-}$, they necessarily have $J^{PC}=1^{--}$. However, all of the $1^{--}$ charmonium states have already been accounted for.

This difficulty was compounded when BABAR's search for an accompanying $Y(4260)\rightarrow \psi(2S)\pi^{+}\pi^{-}$ decay with 298 fb$^{-1}$ of data turned up a structure at a higher mass that is incompatible with the $Y(4260)$. This new state was found to have a mass of $m_{Y}=4324\pm24$ MeV$/c^{2}$ and a width of $\Gamma(Y)=172\pm33$ MeV \cite{y4320_babar}. Belle confirmed this discovery on 670 fb$^{-1}$ of data, measuring $m_{Y}=4361\pm9\pm9$ MeV$/c^{2}$ with a width of $\Gamma(Y)=74\pm15\pm10$ MeV, while finding further evidence for a higher resonance with a mass of $m_{Y}=4664\pm11\pm5$ MeV$/c^{2}$ and width of $\Gamma(Y)=48\pm15\pm3$ MeV \cite{y4320_belle}. These findings now overpopulate $1^{--}$ by four states, making it impossible to explain these resonances within the charmonium model.

\section{Conclusion}
The charmonium model has had great success, but recent experimental results from the $B$-factories is challenging our understanding of the strong force. It is clear that the $X(3872)$ is not a charmonium state; it is likely a $D^{0}\overline{D}^{*0}$ molecule. The nature of the ISR-produced $1^{--}$ states is unclear. Going beyond the charmonium model, lattice QCD and NRQCD will begin to take the lead in the search for a theoretical interpretation. On the experimental front, the BABAR, Belle, and CLEO experiments will remain operational through 2008, followed by the upgraded BES-III thereafter. In the longer term, a Super $B$-factory collaboration offers the possibility of more than an order of magnitude increase in data. This is indeed a very exciting time in the field of quarkonium physics.

\end{document}